\documentclass[12pt]{article}%
\usepackage{amsmath,amssymb,amsthm}
\newtheorem{theorem}{Theorem}
\newtheorem{proposition}{Proposition}
\newtheorem{lemma}[theorem]{Lemma}
\newtheorem{corollary}[theorem]{Corollary}
\begin{document}

\title{Metric Clifford Algebra\thanks{published: \emph{Advances in Applied Clifford
Algebras} \textbf{11}(S3), 49-68 (2001).}}
\author{V. V. Fern\'{a}ndez$^{1}\thanks{e-mail: vvf@ime.unicamp.br}$, A. M.
Moya$^1\thanks{e-mail: moya@ime.unicamp.br}$ and W. A. Rodrigues
Jr.$^1,2\thanks{e-mail: walrod@ime.unicamp.br or walrod@mpc.com.br}$\\$\hspace{-0.5cm}^{1}$ Institute of Mathematics, Statistics and Scientific Computation\\IMECC-UNICAMP CP 6065\\13083-970 Campinas-SP, Brazil\\$^{2}$ Department of Mathematical Sciences, University of Liverpool\\Liverpool, L69 3BX, UK}
\date{10/30/2001}
\maketitle

\begin{abstract}
In this paper we introduce the concept of metric Clifford algebra
$\mathcal{C\ell}(V,g)$ for a $n$-dimensional real vector space $V$ endowed
with a metric extensor $g$ whose signature is $(p,q)$, with $p+q=n$. The
metric Clifford product on $\mathcal{C\ell}(V,g)$ appears as a well-defined
\emph{deformation }(induced by $g$) of an euclidean Clifford product on
$\mathcal{C\ell}(V)$. Associated with the metric extensor $g,$ there is a
gauge metric extensor $h$ which codifies all the geometric information just
contained in $g.$ The precise form of such $h$ is here determined. Moreover,
we present and give a proof of the so-called \emph{golden formula,} which is
important in many applications that naturally appear in ours studies of
multivector functions, and differential geometry and theoretical physics.

\end{abstract}
\tableofcontents

\section{Introduction}

This is paper IV of a series of seven. Here, we introduce the
concept of a metric Clifford algebra for a $n$-dimensional real
vector space $V$ endowed with a metric extensor $g$ of an
arbitrary signature $(p,q),$ with $p+q=n.$ The novelty, regarding
previous presentations of the subject as, e.g., in
(\cite{1},\cite{3}), is that a metric Clifford product appears as
a well-defined \emph{deformation} of an euclidean Clifford
product. More important, we show that associated to any metric
extensor $g$ there is a gauge
metric extensor $h$ (defined `modulus' a gauge) such that $g=h^{\dagger}%
\circ\eta\circ h,$ where $\eta$ is a standard orthogonal metric
extensor over $V$ with the same signature as $g.$ This theorem for
the decomposition of $g$ (which is somewhat analogous to
Silvester's theorem will play a fundamental role in the intrinsic
formulation of the differential geometry on smooth manifolds. The
paper ends with the proof of the so-called \emph{golden formula},
which will show worth to deserve that name. We introduce also the
concepts of standard and metric Hodge (star) operators, and find a
formula connecting them.

\section{Metric Clifford Algebra of Multivectors}

\subsection{Metric Scalar Product}

Let us consider $\bigwedge V$ endowed with the eulidean scalar product
associated to any fixed basis for $V$ denoted by $\{b_{k}\}$, as in previous
papers of this series (\cite{4}-\cite{5}), i.e., the $b$-scalar product on
$\bigwedge V$.

Now, take a $(1,1)$-extensor over $V,$ say $g,$ adjoint symmetric
($g=g^{\dagger}$) and non-degenerate ($\det[g]\neq0$). It will be called a
\emph{metric extensor }over $V.$

We can define another scalar product of multivectors $X,$ $Y\in\bigwedge V$
by
\begin{equation}
X\underset{g}{\cdot}Y=\underline{g}(X)\cdot Y, \label{4.1}%
\end{equation}
where $\underline{g}$ is the\emph{\ extended }of $g.$ It will be called a
\emph{metric scalar product} generated by $g.$ Or, $g$\emph{-scalar product},
for short.

As we can see, this scalar product is a well-defined scalar product on
$\bigwedge V$. It is symmetric, satisfies the distributive laws, has the mixed
associativity property and is non-degenerate, i.e., $X\underset{g}{\cdot}Y=0$
for all $X,$ implies $Y=0.$

All the properties just mentioned above are immediate consequences of the
corresponding ones for the $b$-scalar product. But, a $g$-scalar product is
not necessarily \emph{positive definite}.

We present now some of the most important properties of the $g$-scalar product
of multivectors.\medskip

\textbf{g1} For any $\alpha,\beta\in\mathbb{R}$%
\begin{equation}
\alpha\underset{g}{\cdot}\beta=\alpha\beta\text{ (real product).} \label{4.2}%
\end{equation}

\textbf{g2} For any $X_{j}\in\bigwedge^{j}V$ and $Y_{k}\in\bigwedge^{k}V$
\begin{equation}
X_{j}\underset{g}{\cdot}Y_{k}=0,\text{ if }j\neq k. \label{4.3}%
\end{equation}

\textbf{g3} For any simple $k$-vectors $v_{1}\wedge\ldots v_{k}\in
\bigwedge^{k}V$ and $w_{1}\wedge\ldots w_{k}\in\bigwedge^{k}V$
\begin{equation}
(v_{1}\wedge\ldots v_{k})\underset{g}{\cdot}(w_{1}\wedge\ldots w_{k}%
)=\det\left[
\begin{array}
[c]{ccc}%
v_{1}\underset{g}{\cdot}w_{1} & \ldots & v_{1}\underset{g}{\cdot}w_{k}\\
\ldots & \ldots & \ldots\\
v_{k}\underset{g}{\cdot}w_{1} & \ldots & v_{k}\underset{g}{\cdot}w_{k}%
\end{array}
\right]  , \label{4.4}%
\end{equation}
where (as before) by $\det\left[  v_{p}\underset{g}{\cdot}w_{q}\right]  $ we
denote the classical $k\times k$ determinant.

Eqs.(\ref{4.2}-\ref{4.4}) follow without difficulties from the corresponding
properties of the $b$-scalar product, by taking into account that
$\underline{g}(\alpha)=\alpha$ with $\alpha\in\mathbb{R},$ the
grade-preserving property of the outermorphisms (i.e., of the extension
operator \cite{5}), and that $\underline{g}(v_{1}\wedge\ldots v_{k}%
)=g(v_{1})\wedge\ldots g(v_{k})$ with $v_{1},\ldots,v_{k}\in V.$

\textbf{g4} For any $X,Y\in\bigwedge V$%
\begin{align}
\widehat{X}\underset{g}{\cdot}Y  &  =X\underset{g}{\cdot}\widehat
{Y},\label{4.4a}\\
\widetilde{X}\underset{g}{\cdot}Y  &  =X\underset{g}{\cdot}\widetilde{Y}.
\label{4.4b}%
\end{align}

The proof is immediate and left to the reader. Hint: take into account that
$\widehat{\underline{g}(X)}=\underline{g}(\widehat{X})$ and $\widetilde
{\underline{g}(X)}=\underline{g}(\widetilde{X})$ with $X\in\bigwedge V.$
\subsection{Metric Reciprocal Bases}

Let $(\{e_{k}\},\{e^{k}\})$ be an arbitrary pair of $b$-reciprocal bases of
$V,$ i,e., $e_{k}\cdot e^{l}=\delta_{k}^{l}.$

\begin{theorem}
Take an invertible $(1,1)$-extensor over $V,$ say $f.$ We can construct two
bases for $V,$ say $\{E_{k}\}$ and $\{E^{k}\},$ by the following formulas
\begin{align}
E_{k}  &  =f(e_{k}),\label{4.5a}\\
E^{k}  &  =g^{-1}\circ f^{*}(e^{k})\text{ for each }k=1,\ldots,n. \label{4.5b}%
\end{align}
These bases satisfy the metric scalar product conditions
\begin{equation}
E_{k}\underset{g}{\cdot}E^{l}=\delta_{k}^{l}. \label{4.5c}%
\end{equation}
Reciprocally, given two arbitrary bases $\{E_{k}\}$ and $\{E^{k}\}$ which
satisfy eq.(\ref{4.5c}), there exists an unique invertible $(1,1)$-extensor
$f$ such that the eqs.(\ref{4.5a}) and (\ref{4.5b}) hold.
\end{theorem}

\begin{proof}
Since $\{e_{k}\}$ and $\{e^{k}\}$ are bases for $V$ and, $f$ and $g$ are
invertible $(1,1)$-extensors over $V$, it follows that the $n$ vectors
$E_{1},\ldots,E_{n}\in V$ and the $n$ vectors $E^{1},\ldots,E^{n}\in V$ must
also determine two well-defined bases for $V$.

Now, a straightforward calculation gives
\[
E_{k}\underset{g}{\cdot}E^{l}=g\circ f(e_{k})\cdot g^{-1}\circ f^{*}%
(e^{l})=e_{k}\cdot f^{\dagger}\circ g\circ g^{-1}\circ f^{*}(e^{l})=e_{k}\cdot
e^{l}=\delta_{k}^{l},
\]
and the first statement follows.

Now, $\{e_{k}\}$ and $\{e^{k}\}$ are bases for $V,$ and $\{E_{k}\}$ and
$\{E^{k}\}$ are supposed to be also bases for $V.$ Then, there must exist
exactly two invertible $(1,1)$-extensors over $V,$ say $f_{1}$ and $f_{2},$
such that
\begin{align*}
E_{k}  &  =f_{1}(e_{k}),\\
E^{k}  &  =f_{2}(e^{k})\text{ for each }k=1,\ldots,n.
\end{align*}

It is easy to check that $f_{1}$ and $f_{2}$ are given by
\begin{align*}
f_{1}(v)  &  =(e^{s}\cdot v)E_{s},\\
f_{2}(v)  &  =(e_{s}\cdot v)E^{s}.
\end{align*}

But, using eq.(\ref{4.5c}) we have
\[
f_{1}(e_{k})\underset{g}{\cdot}f_{2}(e^{l})=\delta_{k}^{l}\Rightarrow
e_{k}\cdot f_{1}^{\dagger}\circ g\circ f_{2}(e^{l})=\delta_{k}^{l}\Rightarrow
f_{1}^{\dagger}\circ g\circ f_{2}(e^{l})=e^{l},
\]
for each $l=1,\ldots,n$. Thus, $f_{1}^{\dagger}\circ g\circ f_{2}=i_{V}$.

Then, choosing $f_{1}=f$ and $f_{2}=g^{-1}\circ f^{*},$ the second statement
follows.
\end{proof}

Two bases $\{E_{k}\}$ and $\{E^{k}\}$ satisfying $E_{k}\underset{g}{\cdot
}E^{l}=\delta_{k}^{l}$ are said to be a pair of \emph{metric reciprocal
bases,} and we say that $\{E^{k}\}$ is the \emph{metric reciprocal basis} of
$\{E_{k}\}$.

We end this section presenting two interesting and useful formulas for the
expansion of multivectors in terms of a $g$-scalar product.

\begin{proposition}
Let $(\{E_{k}\},\{E^{k}\})$ be any pair of metric reciprocal bases for $V,$
i.e., $E_{k}\underset{g}{\cdot}E^{l}=\delta_{k}^{l}.$ We have the following
two expansion formulas. For all $X\in\bigwedge V$
\begin{equation}
X=X\underset{g}{\cdot}1+\overset{n}{\underset{k=1}{\sum}}\frac{1}%
{k!}X\underset{g}{\cdot}(E^{j_{1}}\wedge\ldots E^{j_{k}})(E_{j_{1}}%
\wedge\ldots E_{j_{k}}) \label{4.6a}%
\end{equation}
and
\begin{equation}
X=X\underset{g}{\cdot}1+\overset{n}{\underset{k=1}{\sum}}\frac{1}%
{k!}X\underset{g}{\cdot}(E_{j_{1}}\wedge\ldots E_{j_{k}})(E^{j_{1}}%
\wedge\ldots E^{j_{k}}). \label{4.6b}%
\end{equation}

\end{proposition}

The proof is left to the reader. (Hint: use eq.(\ref{4.1}), eq.(\ref{4.5a}),
eq.(\ref{4.5b}) and some of the properties of extension operator, and take
also into account the expansion formula for multivectors in the euclidean
Clifford algebra as defined in \cite{4} (paper I of this series) .

\subsection{Metric Interior Algebras}

We define now the \emph{metric} \emph{left }and \emph{right} \emph{contracted
products }of multivectors $X,$ $Y\in\bigwedge V,$ denoted respectively by
$\underset{g}{\lrcorner}$ and $\underset{g}{\llcorner}, $
\begin{align}
X\underset{g}{\lrcorner}Y  &  =\underline{g}(X)\lrcorner Y,\label{4.7a}\\
X\underset{g}{\llcorner}Y  &  =X\llcorner\underline{g}(Y). \label{4.7b}%
\end{align}
When no confusion arises we call $\underset{g}{\lrcorner}$ and $\underset
{g}{\llcorner}$ the $g$\emph{-contracted products}, for short.

These $g$-contracted products $\underset{g}{\lrcorner}$ and $\underset
{g}{\llcorner}$ are internal laws on $\bigwedge V.$ Both of them satisfy the
distributive laws (on the left and on the right) but they are not associative products.

The vector space $\bigwedge V$ endowed with the $g$-contracted product either
$\underset{g}{\lrcorner}$ or $\underset{g}{\llcorner}$ is a non-associative
algebra. They are called \emph{metric interior algebras of multivectors}. Or,
$g$\emph{-interior algebras}, for short.

We present now some of the basic properties of the metric interior
algebras.\medskip

\textbf{mi1} For any $\alpha,\beta\in\mathbb{R}$ and $X\in\bigwedge V$%
\begin{align}
\alpha\underset{g}{\lrcorner}\beta &  =\alpha\underset{g}{\llcorner}%
\beta=\alpha\beta\text{ (real product),}\label{4.7c}\\
\alpha\underset{g}{\lrcorner}X  &  =X\underset{g}{\llcorner}\alpha=\alpha
X\text{ (multiplication by scalars).} \label{4.7d}%
\end{align}

\textbf{mi2} For any $X_{j}\in\bigwedge^{j}V$ and $Y_{k}\in\bigwedge^{k}V$
with $j\leq k$
\begin{equation}
X_{j}\underset{g}{\lrcorner}Y_{k}=(-1)^{j(k-j)}Y_{k}\underset{g}{\llcorner
}X_{j}. \label{4.7e}%
\end{equation}

\textbf{mi3} For any $X_{j}\in\bigwedge^{j}V$ and $Y_{k}\in\bigwedge^{k}V$
\begin{align}
X_{j}\underset{g}{\lrcorner}Y_{k}  &  =0,\text{ if }j>k,\label{4.7f}\\
X_{j}\underset{g}{\llcorner}Y_{k}  &  =0,\text{ if }j<k. \label{4.7g}%
\end{align}

\textbf{mi4} For any $X_{k},Y_{k}\in\bigwedge^{k}V$
\begin{equation}
X_{k}\underset{g}{\lrcorner}Y_{k}=X_{k}\underset{g}{\llcorner}Y_{k}%
=\widetilde{X_{k}}\underset{g}{\cdot}Y_{k}=X_{k}\underset{g}{\cdot}%
\widetilde{Y_{k}}. \label{4.7h}%
\end{equation}

\textbf{mi5} For any $v\in V$ and $X,Y\in\bigwedge V$
\begin{equation}
v\underset{g}{\lrcorner}(X\wedge Y)=(v\underset{g}{\lrcorner}X)\wedge
Y+\overline{X}\wedge(v\underset{g}{\lrcorner}Y). \label{4.7i}%
\end{equation}

All these properties easily follow from the corresponding properties of the
euclidean interior algebras, once we take into account the properties of
extension operator \cite{4}.

\begin{proposition}
For all $X,Y,Z\in\bigwedge V$ it holds
\begin{align}
(X\underset{g}{\lrcorner}Y)\underset{g}{\cdot}Z  &  =Y\underset{g}{\cdot
}(\widetilde{X}\wedge Z),\label{4.7j}\\
(X\underset{g}{\llcorner}Y)\underset{g}{\cdot}Z  &  =X\underset{g}{\cdot
}(Z\wedge\widetilde{Y}). \label{4.7k}%
\end{align}

\end{proposition}

These properties are completely equivalent to the definitions of the right and
left contracted products given in eqs.(\ref{4.7a}-\ref{4.7b}), and can be
proved without difficulties by using the properties of extension operator
\cite{4}.

\begin{proposition}
For all $X,Y,Z\in\Lambda V$ it holds
\begin{align}
X\underset{g}{\lrcorner}(Y\underset{g}{\lrcorner}Z)  &  =(X\wedge
Y)\underset{g}{\lrcorner}Z,\label{4.7l}\\
(X\underset{g}{\llcorner}Y)\underset{g}{\llcorner}Z  &  =X\underset
{g}{\llcorner}(Y\wedge Z). \label{4.7m}%
\end{align}

\end{proposition}%

\begin{proof}
We prove only the first statement. Take $X,Y,Z\in\bigwedge V$. Using the
multivector identity $A\lrcorner(B\lrcorner C)=(A\wedge B)\lrcorner C$ and a
property of extension operator, we have
\begin{align*}
X\underset{g}{\lrcorner}(Y\underset{g}{\lrcorner}Z)  &  =\underline
{g}(X)\lrcorner(\underline{g}(Y)\lrcorner Z)=(\underline{g}(X)\wedge
\underline{g}(Y))\lrcorner Z\\
&  =\underline{g}(X\wedge Y)\lrcorner Z=(X\wedge Y)\underset{g}{\lrcorner}Z.
\end{align*}
\end{proof}

\subsection{Metric Clifford Algebra}

We define a \emph{metric Clifford product} of $X,Y\in\bigwedge V$ associated
to $g$ by the following axioms:\medskip

\textbf{A1} For all $\alpha\in\mathbb{R}$ and $X\in\bigwedge V$
\[
\alpha\underset{g}{}X=X\alpha\text{ equals multiplication of multivector
}X\text{ by scalar }\alpha.
\]

\textbf{A2} For all $v\in V$ and $X\in\bigwedge V$
\[
v\underset{g}{}X=v\underset{g}{\lrcorner}X+v\wedge X\text{ and }X\underset
{g}{}v=X\underset{g}{\llcorner}v+X\wedge v.
\]

\textbf{A3} For all $X,Y,Z\in\bigwedge V$
\[
X\underset{g}{}(Y\underset{g}{}Z)=(X\underset{g}{}Y)\underset{g}{}Z.
\]

This metric Clifford product is an internal law on $\bigwedge V.$ It is
associative (by the axiom A3) and satisfies the distributive laws (on the left
and on the right) which follow from the corresponding distributive laws of the
euclidean contracted and exterior products \cite{4}.

$\bigwedge V$ endowed with this metric Clifford product is an associative
algebra. It will be called a metric \emph{Clifford algebra of multivectors}
generated by $g$, or simply, $g$\emph{-Clifford algebra. }It will be denoted
by $\mathcal{C}\ell(V,g).$

We present now some of the most basic properties which hold in $\mathcal{C}%
\ell(V,g)$.\medskip

\textbf{clg1} For any $v\in V$ and $X\in\bigwedge V$%
\begin{align}
v\underset{g}{\lrcorner}X  &  =\frac{1}{2}(v\underset{g}{}X-\widehat
{X}\underset{g}{}v)\text{ and }X\underset{g}{\llcorner}v=\frac{1}%
{2}(X\underset{g}{}v-v\underset{g}{}\widehat{X}),\label{4.8a}\\
v\wedge X  &  =\frac{1}{2}(v\underset{g}{}X+\widehat{X}\underset{g}{}v)\text{
and }X\wedge v=\frac{1}{2}(X\underset{g}{}v+v\underset{g}{}\widehat{X}).
\label{4.8b}%
\end{align}

\textbf{clg2} For any $X,Y\in\bigwedge V$%
\begin{equation}
X\underset{g}{\cdot}Y=\left\langle \widetilde{X}\underset{g}{}Y\right\rangle
_{0}=\left\langle X\underset{g}{}\widetilde{Y}\right\rangle _{0}. \label{4.8c}%
\end{equation}

\textbf{clg3} For $X,Y,Z\in\bigwedge V$%
\begin{align}
(X\underset{g}{}Y)\underset{g}{\cdot}Z  &  =Y\underset{g}{\cdot}(\widetilde
{X}\underset{g}{}Z)=X\underset{g}{\cdot}(Z\underset{g}{}\widetilde
{Y}),\label{4.8d}\\
X\underset{g}{\cdot}(Y\underset{g}{}Z)  &  =(\widetilde{Y}\underset{g}%
{}X)\underset{g}{\cdot}Z=(X\underset{g}{}\widetilde{Z})\underset{g}{\cdot}Y.
\label{4.8e}%
\end{align}

\textbf{clg4} For any $X,Y\in\bigwedge V$%
\begin{align}
\widehat{X\underset{g}{}Y}  &  =\widehat{X}\underset{g}{}\widehat
{Y},\label{4.8f}\\
\widetilde{X\underset{g}{}Y}  &  =\widetilde{Y}\underset{g}{}\widetilde{X}.
\label{4.8g}%
\end{align}

\textbf{clg5} Let $I\in\bigwedge^{n}V,$ then for any $v\in V$ and
$X\in\bigwedge V$%
\begin{equation}
I\underset{g}{}(v\wedge X)=(-1)^{n-1}v\underset{g}{\lrcorner}(I\underset{g}%
{}X). \label{4.8h}%
\end{equation}
Eq.(\ref{4.8h}) will be called the metric duality identity, or $g$-duality
identity, for short.

\section{Eigenvalues and Eigenvectors}

Let $t$ be a $(1,1)$-extensor over $V.$ A scalar $\lambda\in\mathbb{R}$ and a
non-zero vector $v\in V$ are said to be an eigenvalue and an eigenvector of
$t,$ respectively, if and only if
\begin{equation}
t(v)=\lambda v. \label{4.9}%
\end{equation}

We say that $\lambda$ and $v$ are naturally to be associated to each other.
This means that, if $\lambda\in\mathbb{R}$ is an eigenvalue of $t,$ then there
is some non-zero $v\in V$ (the associated eigenvector of $t$) such that
eq.(\ref{4.9}) holds, and if a non-zero $v\in V$ is an eigenvector of $t,$
then there is some $\lambda\in\mathbb{R}$ (the associated eigenvalue of $t $)
such that eq.(\ref{4.9}) is satisfied.

A scalar $\lambda\in\mathbb{R}$ is an eigenvalue of $t$ if and only if it
satisfies the following algebraic equation of degree $n$
\begin{equation}
\det[\lambda i_{V}-t]=0, \label{4.9a}%
\end{equation}
where $i_{V}\in ext_{1}^{1}(V)$ is the known identity $(1,1)$-extensor over
$V.$

\begin{theorem}
For any adjoint symmetric $(1,1)$-extensor $s,$ i.e., $s=s^{\dagger},$ there
exists a set of $n$ eigenvectors of $s$ which is a $b$-orthonormal basis for
$V.$ This means that there are exactly $n$ linearly independent non-zero
vectors $v_{1},\ldots,v_{n}\in V$ and $n$ scalars $\lambda_{1},\ldots
,\lambda_{n}\in\mathbb{R}$ such that
\[
s(v_{k})=\lambda_{k}v_{k},\text{ for each }k=1,\ldots,n
\]
and $\{v_{k}\}$ is a basis for $V$ which satisfies $v_{j}\cdot v_{k}%
=\delta_{jk}.$
\end{theorem}

\begin{corollary}
All eigenvalues of a metric extensor $g$ (i.e., $g\in ext_{1}^{1}(V),$
$g=g^{\dagger}$ and $\det[g]\neq0$) are non-zero real numbers.
\end{corollary}

\begin{proof}
We first calculate $\det[g]$ by using $\{v_{k}\},$ taking into account the
eigenvalue equation of $g$ and recalling that, in this case, the euclidean
reciprocal vectors $v^{k}$ are equal to the vectors $v_{k},$%
\begin{align}
\det[g]  &  =(g(v_{1})\wedge\ldots g(v_{n}))\cdot(v^{1}\wedge\ldots
v^{n})\nonumber\\
&  =(\lambda_{1}v_{1}\wedge\ldots\lambda_{n}v_{n})\cdot(v_{1}\wedge\ldots
v_{1})\nonumber\\
&  =\lambda_{1}\ldots\lambda_{n}(v_{1}\wedge\ldots v_{1})\cdot(v_{1}%
\wedge\ldots v_{1})\nonumber\\
\det[g]  &  =\lambda_{1}\ldots\lambda_{n}. \label{4.9b}%
\end{align}
Since $\det[g]\neq0,$ all $\lambda_{1},\ldots,\lambda_{n}$ must be non-zero
real numbers.
\end{proof}

The integer number $s=p-q$, where $p,q$ are non negative integer numbers,
respectively the numbers of positive and negative eigenvalues of $t$ and
$p+q=n,$ is called the signature of $t$. We already have used (and will
continue to do so) the usual convention of physicists and denote the signature
of $g$ by the pair $(p,q)$.

\section{Gauge Metric Extensor}

\begin{lemma}
Any $b$-orthogonal symmetric $(1,1)$-extensor over $V,$ say $\sigma,$ (i.e.,
$\sigma=\sigma^{*}$ and $\sigma=\sigma^{\dagger}$) can only have eigenvalues
$\pm1$.
\end{lemma}

\begin{proof}
If $\lambda\in R$ is an eigenvalue of $\sigma,$ there is an non-zero $v\in V,$
the associated eigenvector of $\sigma,$ such that $\sigma(v)=\lambda v. $ And,
the orthogonality and symmetry of $\sigma$ yield $\sigma^{2}=i_{V}.$

Thus, we have that $v=\lambda^{2}v.$ Since $v\neq0,$ it follows that
$1-\lambda^{2}=0,$ i.e., $\lambda=\pm1.$
\end{proof}

\begin{lemma}
Let $\{b_{k}\}$ be the fiducial basis for $V$ (i.e., $b_{j}\underset{b}{\cdot
}b_{k}=\delta_{jk}$)$.$ We can construct a \emph{fiducial }$b$-orthogonal
metric extensor over $V,$ say $\eta\in ext_{1}^{1}(V),$ (i.e., $\eta=\eta^{*}$
and $\eta=\eta^{\dagger},$ $\det[\eta]\neq0$) with signature $(p,q)$ and which
$b$-orthonormal basis of $V,$ made of the eigenvectors of $\eta,$ is exactly
$\{b_{k}\}.$

Such a $(1,1)$-extensor over $V$ is given by
\begin{equation}
\eta(v)=\overset{p}{\underset{j=1}{\sum}}(v\cdot b_{j})b_{j}-\overset
{p+q}{\underset{j=p+1}{\sum}}(v\cdot b_{j})b_{j}. \label{4.10}%
\end{equation}

\end{lemma}%

\begin{proof}
We first shall prove that $\eta$ has $p$ eigenvalues $+1$ with associated
eigenvectors $b_{1},\ldots,b_{p}$ and $q$ eigenvalues $-1$ with associated
eigenvectors $b_{p+1},\ldots,b_{p+q}.$

Take $b_{k}$ with $k=1,\ldots,p$ we have
\[
\eta(b_{k})=\overset{p}{\underset{j=1}{\sum}}(b_{k}\cdot b_{j})b_{j}%
-\overset{p+q}{\underset{j=p+1}{\sum}}(b_{k}\cdot b_{j})b_{j}=\overset
{p}{\underset{j=1}{\sum}}\delta_{kj}b_{j}-\overset{p+q}{\underset{j=p+1}{\sum
}}0b_{j}=b_{k},
\]
and, for $b_{k}$ with $k=p+1,\ldots,p+q,$ it yields
\[
\eta(b_{k})=\overset{p}{\underset{j=1}{\sum}}(b_{k}\cdot b_{j})b_{j}%
-\overset{p+q}{\underset{j=p+1}{\sum}}(b_{k}\cdot b_{j})b_{j}=\overset
{p}{\underset{j=1}{\sum}}0b_{j}-\overset{p+q}{\underset{j=p+1}{\sum}}%
\delta_{kj}b_{j}=-b_{k}.
\]
Then, we have
\begin{equation}
\eta(b_{k})=\left\{
\begin{array}
[c]{ll}%
b_{k}, & k=1,\ldots,p\\
-b_{k}, & k=p+1,\ldots,p+q
\end{array}
\right.  . \label{4.10bis}%
\end{equation}

Now, we shall prove that $\eta$ as defined above is a metric extensor over
$V,$ i.e., $\eta=\eta^{\dagger}$ and $\det[\eta]\neq0.$

Take $v,w\in V$ then
\begin{align*}
\eta^{\dagger}(v)\cdot w  &  =v\cdot\eta(w)=v\cdot(\overset{p}{\underset
{j=1}{\sum}}(w\cdot b_{j})b_{j}-\overset{p+q}{\underset{j=p+1}{\sum}}(w\cdot
b_{j})b_{j})\\
&  =\overset{p}{\underset{j=1}{\sum}}(v\cdot b_{j})(w\cdot b_{j}%
)-\overset{p+q}{\underset{j=p+1}{\sum}}(v\cdot b_{j})(w\cdot b_{j})\\
&  =(\overset{p}{\underset{j=1}{\sum}}(v\cdot b_{j})b_{j}-\overset
{p+q}{\underset{j=p+1}{\sum}}(v\cdot b_{j})b_{j})\cdot w=\eta(v)\cdot w,
\end{align*}
i.e., $\eta^{\dagger}=\eta$.

We calculate the determinant of $\eta$ by using the fundamental formula with
$\{b_{k}\},$ i.e., $\det[\eta]=\eta(b_{1})\wedge\ldots\eta(b_{n})\cdot
(b^{1}\wedge\ldots b^{n}).$ Recall that, in this case, the $b$-reciprocal
basis vectors $b^{k}$ coincide with $b_{k}$ for $k=1,\ldots,n.$%
\begin{align*}
&  \det[\eta]\\
&  =(\eta(b_{1})\wedge\ldots\eta(b_{p})\wedge\eta(b_{p+1})\wedge\ldots
\eta(b_{p+q}))\cdot(b_{1}\wedge\ldots b_{p}\wedge b_{p+1}\wedge\ldots
b_{p+q})\\
&  =(b_{1}\wedge\ldots b_{p}\wedge(-1)^{q}b_{p+1}\wedge\ldots b_{p+q}%
)\cdot(b_{1}\wedge\ldots b_{p}\wedge b_{p+1}\wedge\ldots b_{p+q}),
\end{align*}
i.e., $\det[\eta]=(-1)^{q}.$

Next, we shall prove that $\eta^{2}=i_{V},$ i.e., $\eta^{-1}=\eta$.

Take $v\in V$ then
\begin{align*}
\eta\circ\eta(v)  &  =\overset{n}{\underset{k=1}{\sum}}(v\cdot b_{k})\eta
\circ\eta(b_{k})\\
&  =\overset{n}{\underset{k=1}{\sum}}(v\cdot b_{k})\left\{
\begin{array}
[c]{ll}%
\eta(b_{k}) & k=1,\ldots,p\\
-\eta(b_{k}) & k=p+1,\ldots,p+q
\end{array}
\right. \\
&  =\overset{n}{\underset{k=1}{\sum}}(v\cdot b_{k})b_{k}=v,
\end{align*}
i.e., $\eta^{2}=i_{V}.$
\end{proof}

This lemma allows us to construct, associated to the fiducial basis
$\{b_{k}\},$ a fiducial $b$-orthogonal metric extensor $\eta\in ext_{1}%
^{1}(V)$ with signature $(p,q)$. Such a $(1,1)$-extensor over $V$ has $p$
eigenvalues $+1$ and $q$ eigenvalues $-1,$ and their corresponding associated
eigenvectors are the vectors of $\{b_{k}\}.$

\begin{theorem}
For any metric extensor $g\in ext_{1}^{1}(V)$ whose signature is $(p,q),$
there exists an invertible extensor $h\in ext_{1}^{1}(V)$ such that
\begin{equation}
g=h^{\dagger}\circ\eta\circ h, \label{4.11}%
\end{equation}
where $\eta\in ext_{1}^{1}(V)$ is just the fiducial $b$-orthogonal metric
extensor with signature $(p,q),$ as considered in eq.(\ref{4.10}).

Such a $(1,1)$-extensor over $V$ is given by
\begin{equation}
h(a)=\overset{n}{\underset{j=1}{\sum}}\sqrt{\left|  \lambda_{j}\right|
}(a\cdot v_{j})b_{j}, \label{4.11a}%
\end{equation}
where $\lambda_{1},\ldots\lambda_{n}\in\mathbb{R}$ are the eigenvalues of $g $
and $v_{1},\ldots,v_{n}\in V$ are the corresponding associated eigenvectors of
$g.$
\end{theorem}

\begin{proof}
First we need calculate the adjoint extensor of $h.$

Take $a,b\in V$ then
\begin{align*}
h^{\dagger}(a)\cdot b  &  =a\cdot h(b)=a\cdot(\overset{n}{\underset{j=1}{\sum
}}\sqrt{\left|  \lambda_{j}\right|  }(b\cdot v_{j})b_{j})\\
&  =(\overset{n}{\underset{j=1}{\sum}}\sqrt{\left|  \lambda_{j}\right|
}(a\cdot b_{j})v_{j})\cdot b,
\end{align*}
i.e., $h^{\dagger}(a)=\overset{n}{\underset{j=1}{\sum}}\sqrt{\left|
\lambda_{j}\right|  }(a\cdot b_{j})v_{j}.$

Now, let $a\in V.$ A straightforward calculation yields
\begin{align*}
h^{\dagger}\circ\eta\circ h(a)  &  =\overset{n}{\underset{j=1}{\sum}}%
\overset{n}{\underset{k=1}{\sum}}\sqrt{\left|  \lambda_{j}\lambda_{k}\right|
}\eta(b_{j})\cdot b_{k}(a\cdot v_{j})v_{k}\\
&  =\overset{p}{\underset{j=1}{\sum}}\overset{p}{\underset{k=1}{\sum}}%
\sqrt{\left|  \lambda_{j}\lambda_{k}\right|  }\eta(b_{j})\cdot b_{k}(a\cdot
v_{j})v_{k}\\
&  +\overset{p}{\underset{j=1}{\sum}}\overset{p+q}{\underset{k=p+1}{\sum}%
}\sqrt{\left|  \lambda_{j}\lambda_{k}\right|  }\eta(b_{j})\cdot b_{k}(a\cdot
v_{j})v_{k}\\
&  +\overset{p+q}{\underset{j=p+1}{\sum}}\overset{p}{\underset{k=1}{\sum}%
}\sqrt{\left|  \lambda_{j}\lambda_{k}\right|  }\eta(b_{j})\cdot b_{k}(a\cdot
v_{j})v_{k}\\
&  +\overset{p+q}{\underset{j=p+1}{\sum}}\overset{p+q}{\underset{k=p+1}{\sum}%
}\sqrt{\left|  \lambda_{j}\lambda_{k}\right|  }\eta(b_{j})\cdot b_{k}(a\cdot
v_{j})v_{k},
\end{align*}
and, by taking into account eq.(\ref{4.10bis}) we have
\begin{align*}
h^{\dagger}\circ\eta\circ h(a)  &  =\overset{p}{\underset{j=1}{\sum}}%
\overset{p}{\underset{k=1}{\sum}}\sqrt{\left|  \lambda_{j}\lambda_{k}\right|
}\delta_{jk}(a\cdot v_{j})v_{k}+0\\
&  +0-\overset{p+q}{\underset{j=p+1}{\sum}}\overset{p+q}{\underset{k=p+1}%
{\sum}}\sqrt{\left|  \lambda_{j}\lambda_{k}\right|  }\delta_{jk}(a\cdot
v_{j})v_{k}\\
&  =\overset{p}{\underset{j=1}{\sum}}\left|  \lambda_{j}\right|  (a\cdot
v_{j})v_{j}-\overset{p+q}{\underset{j=p+1}{\sum}}\left|  \lambda_{j}\right|
(a\cdot v_{j})v_{j}\\
&  =\overset{p}{\underset{j=1}{\sum}}\lambda_{j}(a\cdot v_{j})v_{j}%
+\overset{p+q}{\underset{j=p+1}{\sum}}\lambda_{j}(a\cdot v_{j})v_{j}%
=\overset{n}{\underset{j=1}{\sum}}\lambda_{j}(a\cdot v_{j})v_{j}.
\end{align*}
On the last step we have used that the signature of $g$ is $(p,q),$ i.e., $g$
has $p$ positive eigenvalues and $q$ negative eigenvalues.

And, by using the eigenvalues equation of $g,$ i.e., $g(v_{j})=\lambda
_{j}v_{j}$ for each $j=1,\ldots,n,$ we have
\[
h^{\dagger}\circ\eta\circ h(a)=\overset{n}{\underset{j=1}{\sum}}(a\cdot
v_{j})g(v_{j})=g(\overset{n}{\underset{j=1}{\sum}}(a\cdot v_{j})v_{j})=g(a),
\]
i.e., $h^{\dagger}\circ\eta\circ h=g.$

Finally, since
\[
\det[g]=\det[h^{\dagger}\circ\eta\circ h]=\det[h^{\dagger}]\det[\eta
]\det[h]=\det[\eta]\left.  \det\right.  ^{2}[h]=(-1)^{q}\left.  \det\right.
^{2}[h],
\]
and $\det[g]\neq0,$ then $\det[h]\neq0,$ and so $h$ is an invertible
$(1,1)$-extensor.
\end{proof}

It should be noted that $h$ satisfying eq.(\ref{4.11}) is not \emph{unique}.
If there is some $h\in ext_{1}^{1}(V)$ which satisfies eq.(\ref{4.11}), then
$h^{\prime}\equiv\Lambda\circ h,$ where $\Lambda$ is a $\eta$-orthogonal
$(1,1)$-extensor over $V$ (i.e., $\Lambda^{\dagger}\circ\eta\circ\Lambda=\eta
$)\footnote{As the reader can prove without difficulties, a $\eta$-orthogonal
$(1,1)$-extensor $\Lambda$ preserves the $\eta$-scalar products, i.e., for all
$v,w\in V:\Lambda(v)\underset{\eta}{\cdot}\Lambda(w)=v\underset{\eta}{\cdot
}w.$}, also satisfies eq.(\ref{4.11}).

Indeed, we have $h^{\prime\dagger}\circ\eta\circ h^{\prime}=(\Lambda\circ
h)^{\dagger}\circ\eta\circ\Lambda\circ h=h^{\dagger}\circ\Lambda^{\dagger
}\circ\eta\circ\Lambda\circ h=h^{\dagger}\circ\eta\circ h=g.$

In general, an invertible extensor $h\in ext_{1}^{1}(V)$ which satisfies
eq.(\ref{4.11}) will be said to be \emph{a gauge metric extensor for the
metric extensor }$g.$

\subsection{Gauge Metric Bases}

Let $\{e_{k}\}$ and $\{e^{k}\}$ be two $b$-reciprocal bases to each other for
the vector space $V,$ i.e., $e_{k}\cdot e^{l}=\delta_{k}^{l}.$ Since $h, $ a
gauge metric extensor over $V,$ is non-degenerate, i.e., $\det[h]\neq0$, it
follows that the $n$ vectors $h(e_{1}),\ldots,h(e_{n})\in V$ and the $n $
vectors $h^{*}(e^{1}),\ldots,h^{*}(e^{n})\in V$ will be also well-defined
bases for $V.$

As the reader can easily prove $(\{h(e_{k})\},$ $\{h^{*}(e^{k})\})$ is also a
pair of $b$-reciprocal bases of $V,$ i.e.,
\begin{equation}
h(e_{k})\cdot h^{*}(e^{l})=\delta_{k}^{l}. \label{4.11b}%
\end{equation}

Two other remarkable properties of these bases are:
\begin{align}
h(e_{j})\underset{\eta}{\cdot}h(e_{k})  &  =g(e_{j})\cdot e_{k}\equiv
g_{jk},\label{4.11c}\\
h^{*}(e^{j})\underset{\eta^{-1}}{\cdot}h^{*}(e^{k})  &  =g^{-1}(e^{j})\cdot
e^{k}\equiv g^{jk}. \label{4.11d}%
\end{align}

These bases $\{h(e_{k})\}$ and $\{h^{*}(e^{k})\}$ will be said to be a pair of
\emph{gauge metric bases} for $V.$

\begin{theorem}
Given $n$ non-zero real numbers $\rho_{1},\ldots,\rho_{n}$ and a
$b$-orthogonal $(1,1)$-extensor $l$ over $V$ (i.e., $l=l^{*}$), we can
construct an invertible $(1,1)$-extensor $h$ over $V$ using the following
formula
\begin{equation}
h(v)=\overset{n}{\underset{j=1}{\sum}}\rho_{j}(l(v)\cdot b_{j})b_{j}.
\label{4.12}%
\end{equation}

Then, the $(1,1)$-extensor $g$ over $V$ defined by
\begin{equation}
g=h^{\dagger}\circ\eta\circ h, \label{4.12a}%
\end{equation}
where $\eta\in ext_{1}^{1}(V)$ is just the fiducial $b$-orthogonal metric
extensor over $V$ with signature $(p,q),$ as considered in eq.(\ref{4.10}).

The $p$ positive real numbers $\rho_{1}^{2},\ldots,\rho_{p}^{2}$ are the
eigenvalues of $g$ with the associated eigenvectors $l^{\dagger}(b_{1}%
),\ldots,l^{\dagger}(b_{p})$ of $g,$ and the $q$ negative real numbers
$-\rho_{p+1}^{2},\ldots,-\rho_{p+q}^{2}$ are the eigenvalues of $g $ with the
associated eigenvectors $l^{\dagger}(b_{p+1}),\ldots,l^{\dagger}(b_{p+q})$ of
$g.$

The set of $n$ non-zero vectors $\{l^{\dagger}(b_{1}),\ldots,l^{\dagger}%
(b_{p}),$ $l^{\dagger}(b_{p+1}),\ldots,l^{\dagger}(b_{p+q})\}$ is a
$b$-orthonormal basis of $V,$ made of the eigenvectors of $g.$

The signature of $g$ is also $(p,q).$
\end{theorem}

\begin{proof}
We first must check that $g$ is symmetric and non-degenerate. Using
eq.(\ref{4.12a}) we have
\[
g^{\dagger}=h^{\dagger}\circ\eta^{\dagger}\circ(h^{\dagger})^{\dagger
}=h^{\dagger}\circ\eta\circ h=g,
\]
thus, $g=g^{\dagger},$ i.e., $g$ is symmetric.

We now calculate $\det[g],$
\[
\det[g]=\det[h^{\dagger}]\det[\eta]\det[h]=\det[h]\det[\eta]\det
[h]=(-1)^{q}\left.  \det\right.  ^{2}[h].
\]

But, it is possible to calculate $\det[h]$ by using a trick. We shall evaluate
$\det[h\circ l^{\dagger}]$ in two different ways.

First, using eq.(\ref{4.12}) and the $b$-orthogonality of $l,$ i.e.,
$l^{-1}=l^{\dagger},$ we have
\[
h\circ l^{\dagger}(v)=\overset{n}{\underset{j=1}{\sum}}\rho_{j}(a\cdot
b_{j})b_{j}.
\]

Sencondly, using the fundamental formula for the determinant of a
$(1,1)$-extensor (see [5]) we have,
\begin{align*}
\det[h\circ l^{\dagger}]  &  =(h\circ l^{\dagger}(b_{1})\wedge\ldots h\circ
l^{\dagger}(b_{n}))\cdot(b^{1}\wedge\ldots b^{n})\\
&  =(\rho_{1}b_{1}\wedge\ldots\rho_{n}b_{n})\cdot(b_{1}\wedge\ldots b_{n})\\
&  =\rho_{1}\ldots\rho_{n}.
\end{align*}

Now, taking into account that $\det[l]=\pm1$ and property (d1) of the
determinant (see [5]) we get,
\[
\det[h\circ l^{\dagger}]=\det[h]\det[l^{\dagger}]=\det[h]\det[l]=\pm\det[h].
\]

Thus, we have $\det[h]=\pm\rho_{1}\ldots\rho_{n}$. And, therefore
$\det[g]=(-1)^{q}\rho_{1}^{2}\ldots\rho_{n}^{2}.$ Since $\rho_{1},\ldots
,\rho_{n}$ are non-zero real numbers, $\det[g]\neq0,$ i.e., $g$ is non-degenerate.

The proof of the first statement is then complete.

In order to prove the second statement, related to the eigenvalues and
eigenvectors of $g,$ we shall use the following equations: $h\circ l^{\dagger
}(b_{k})=\rho_{k}b_{k}$ (just used above), $h^{\dagger}(b_{k})=\rho
_{k}l^{\dagger}(b_{k})$ (obtained from eq.(\ref{4.12})) and the eigenvalue
equation of $\eta,$ i.e., $\eta(b_{k})=\left\{
\begin{array}
[c]{ll}%
b_{k}, & k=1,\ldots,p\\
-b_{k}, & k=p+1,\ldots,p+q
\end{array}
\right.  .$

We have
\begin{align*}
g\circ l^{\dagger}(b_{k})  &  =h^{\dagger}\circ\eta\circ h\circ l^{\dagger
}(b_{k})=h^{\dagger}\circ\eta(\rho_{k}b_{k})=\rho_{k}h^{\dagger}\circ
\eta(b_{k})\\
&  =\left\{
\begin{array}
[c]{ll}%
\rho_{k}h^{\dagger}(b_{k}), & k=1,\ldots,p\\
-\rho_{k}h^{\dagger}(b_{k}), & k=p+1,\ldots,p+q
\end{array}
\right. \\
g\circ l^{\dagger}(e_{k})  &  =\left\{
\begin{array}
[c]{ll}%
\rho_{k}^{2}l^{\dagger}(b_{k}), & k=1,\ldots,p\\
-\rho_{k}^{2}l^{\dagger}(b_{k}), & k=p+1,\ldots,p+q
\end{array}
\right.  .
\end{align*}

This establishes the second statement.

>From the euclidean $b$-orthogonality of $l$ it follows easily that the
eigenvectors of $g,$ are $b$-orthonormal. It is also obvious that the
signature of $g$ is also $(p,q).$ Thus, the third and fourth statement are
proved.
\end{proof}

\section{The Golden Formula}

\begin{proposition}
Let $h$ be any gauge operator for $g,$ i.e., $g=h^{\dagger}\circ\eta\circ h,$
and let $\underset{g}{*\text{ }}$mean either $\wedge$ (exterior product),
$\underset{g}{\cdot\text{ }}$($g$-scalar product),$\underset{g}{\text{
}\lrcorner}\underset{g}{\llcorner\text{ }}$($g$-contracted products) or
$\underset{g}{}$ ($g$-Clifford product), and analogously for $\underset{\eta
}{*}$. The $g$-metric products $\underset{g}{*}$ and the $\eta$-metric
products $\underset{\eta}{*}$ are related by the following remarkable formula.
For all $X,Y\in\bigwedge V$
\begin{equation}
\underline{h}(X\underset{g}{*}Y)=\underline{h}(X)\underset{\eta}{*}%
\underline{h}(Y), \label{4.golden}%
\end{equation}
where $\underline{h}$ denotes the extended of $h$. Eq.(\ref{4.golden}) will be
called the \emph{golden} \emph{formula}
\end{proposition}

\begin{proof}

By recalling the fundamental properties for the outermorphism of an operator:
$\underline{t}(X\wedge Y)=\underline{t}(X)\wedge\underline{t}(Y)$ and
$\underline{t}(\alpha)=\alpha,$ we have that the multivector identity above
holds for the exterior product, i.e.,
\begin{equation}
X\wedge Y=\underline{h}^{-1}[\underline{h}(X)\wedge\underline{h}(Y)]
\label{4.g1}%
\end{equation}
and for the $g$-scalar product and the $\eta$-scalar product, i.e.,
\begin{equation}
X\underset{g}{\cdot}Y=\underline{h}^{-1}[\underline{h}(X)\underset{\eta}%
{\cdot}\underline{h}(Y)]. \label{4.g2}%
\end{equation}

By using the multivector identities for an invertible operator: $\underline
{t}^{\dagger}(X)\lrcorner Y=\underline{t}^{-1}[X\lrcorner\underline{t}(Y)]$
and $X\llcorner\underline{t}^{\dagger}(Y)=\underline{t}^{-1}[\underline
{t}(X)\llcorner Y],$ and the gauge equation $g=h^{\dagger}\circ\eta\circ h$ we
can easily prove that the multivector identity above holds for the
$g$-contracted product and the $\eta$-contracted product, i.e.,
\begin{align}
X\underset{g}{\lrcorner}Y  &  =\underline{h}^{-1}[\underline{h}(X)\underset
{\eta}{\lrcorner}\underline{h}(Y)]\label{4.g3.1}\\
X\underset{g}{\llcorner}Y  &  =\underline{h}^{-1}[\underline{h}(X)\underset
{\eta}{\llcorner}\underline{h}(Y)]. \label{4.g3.2}%
\end{align}

To prove eq.(\ref{4.g3.1}) see that we can write
\[
X\underset{g}{\lrcorner}Y=\underline{h^{\dagger}\circ\eta\circ h}(X)\lrcorner
Y=\underline{h}^{-1}[\underline{\eta\circ h}(X)\lrcorner\underline
{h}(Y)]=\underline{h}^{-1}[\underline{h}(X)\underset{\eta}{\lrcorner
}\underline{h}(Y)],
\]
where the definitions of $\underset{g}{\lrcorner}$ and $\underset{\eta
}{\lrcorner}$ have been used. The proof of eq.(\ref{4.g3.2}) is completely
analogous, the definitions of $\underset{g}{\llcorner}$ and $\underset{\eta
}{\llcorner}$ should be used.

In order to prove that the multivector identity above holds for the
$g$-Clifford product and the $\eta$-Clifford product, i.e.,
\begin{equation}
X\underset{g}{}Y=\underline{h}^{-1}[\underline{h}(X)\underset{\eta}%
{}\underline{h}(Y)], \label{4.g4}%
\end{equation}
we first must prove four particular cases of it.

Take $\alpha\in\mathbb{R}$ and $X\in\bigwedge V.$ By using the axioms of the
$g$ and $\eta$ Clifford products: $\alpha\underset{g}{}X=X\underset{g}{}%
\alpha=\alpha X$ and $\alpha\underset{\eta}{}X=X\underset{\eta}{}\alpha=\alpha
X,$ we can write
\[
\alpha\underset{g}{}X=\alpha X=\underline{h}^{-1}[\alpha\underline
{h}(X)]=\underline{h}^{-1}[\alpha\underset{\eta}{}\underline{h}(X)],
\]
i.e.,
\begin{equation}
\alpha\underset{g}{}X=\underline{h}^{-1}[\underline{h}(\alpha)\underset{\eta
}{}\underline{h}(X)]. \label{4.g5}%
\end{equation}
Analogously, we have
\begin{equation}
X\underset{g}{}\alpha=\underline{h}^{-1}[\underline{h}(X)\underset{\eta}%
{}\underline{h}(\alpha)]. \label{4.g6}%
\end{equation}

Take $v\in V$ and $X\in\bigwedge V.$ By using the axioms of the $g$ and $\eta$
Clifford products: $v\underset{g}{}X=v\underset{g}{\lrcorner}X+v\wedge X$ and
$v\underset{\eta}{}X=v\underset{\eta}{\lrcorner}X+v\wedge X,$ and
eqs.(\ref{4.g3.1}) and (\ref{4.g1}) we can write
\[
v\underset{g}{}X=v\underset{g}{\lrcorner}X+v\wedge X=\underline{h}%
^{-1}[h(v)\underset{\eta}{\lrcorner}\underline{h}(X)]+\underline{h}%
^{-1}[h(v)\wedge\underline{h}(X)],
\]
i.e.,
\begin{equation}
v\underset{g}{}X=\underline{h}^{-1}[h(v)\underset{\eta}{}\underline{h}(X)].
\label{4.g7}%
\end{equation}
>From the axioms of the $g$ and $\eta$ Clifford products: $X\underset{g}%
{}v=X\underset{g}{\llcorner}v+X\wedge v$ and $X\underset{\eta}{}%
v=X\underset{\eta}{\llcorner}v+X\wedge v$, and eqs.(\ref{4.g3.2}) and
(\ref{4.g1}) we get
\begin{equation}
X\underset{g}{}v=\underline{h}^{-1}[\underline{h}(X)\underset{\eta}{}h(v)].
\label{4.g8}%
\end{equation}

Take $v_{1},v_{2},\ldots,v_{k}\in V.$ By using $k-1$ times eq.(\ref{4.g7}) we
have indeed
\begin{align}
v_{1}\underset{g}{}v_{2}\underset{g}{\cdots}v_{k}  &  =\underline{h}%
^{-1}[h(v_{1})\underset{\eta}{}\underline{h}(v_{2}\underset{g}{\cdots}%
v_{k})]\nonumber\\
&  =\underline{h}^{-1}[h(v_{1})\underset{\eta}{}h(v_{2})\underset{\eta}%
{\cdots}h(v_{k})],\nonumber\\
v_{1}\underset{g}{}v_{2}\underset{g}{\cdots}v_{k}  &  =\underline{h}%
^{-1}[h(v_{1})\underset{\eta}{}h(v_{2})\underset{\eta}{\cdots}h(v_{k})].
\label{4.g9}%
\end{align}

Take $v_{1},v_{2},\ldots,v_{k}\in V$ and $X\in\bigwedge V.$ By using $k-1$
times eq.(\ref{4.g7}) and eq.(\ref{4.g9}) we have indeed
\begin{align}
(v_{1}\underset{g}{}v_{2}\underset{g}{\cdots}v_{k})\underset{g}{}X  &
=\underline{h}^{-1}[h(v_{1})\underset{\eta}{}\underline{h}(\underset{g}{}%
v_{2}\underset{g}{\cdots}v_{k}\underset{g}{}X)]\nonumber\\
&  =\underline{h}^{-1}[h(v_{1})\underset{\eta}{}h(v_{2})\underset{\eta}%
{\cdots}h(v_{k})\underset{\eta}{}\underline{h}(X)],\nonumber\\
(v_{1}\underset{g}{}v_{2}\underset{g}{\cdots}v_{k})\underset{g}{}X  &
=\underline{h}^{-1}[\underline{h}(v_{1}\underset{g}{}v_{2}\underset{g}{\cdots
}v_{k})\underset{\eta}{}\underline{h}(X)]. \label{4.g10}%
\end{align}

We now can prove the general case of eq.(\ref{4.g4}). We shall use an
expansion formula for multivectors: $X=X^{0}+\underset{k=1}{\overset{n}{\sum}%
}\dfrac{1}{k!}X^{j_{1}\ldots j_{k}}e_{j_{1}}\underset{g}{\cdots}e_{j_{k}},$
where $\{e_{j}\}$ is a basis of $V,$ eq.(\ref{4.g5}) and eq.(\ref{4.g10}). We
can write
\begin{align*}
X\underset{g}{}Y  &  =X^{0}\underset{g}{}Y+\underset{k=1}{\overset{n}{\sum}%
}\frac{1}{k!}X^{j_{1}\ldots j_{k}}(e_{j_{1}}\underset{g}{\cdots}e_{j_{k}%
})\underset{g}{}Y\\
&  =\underline{h}^{-1}[\underline{h}(X^{0})\underset{\eta}{}\underline
{h}(Y)]+\underline{h}^{-1}[\underset{k=1}{\overset{n}{\sum}}\frac{1}%
{k!}X^{j_{1}\ldots j_{k}}\underline{h}(e_{j_{1}}\underset{g}{\cdots}e_{j_{k}%
})\underset{\eta}{}\underline{h}(Y)]\\
&  =\underline{h}^{-1}[\underline{h}(X^{0}+\underset{k=1}{\overset{n}{\sum}%
}\frac{1}{k!}X^{j_{1}\ldots j_{k}}e_{j_{1}}\underset{g}{\cdots}e_{j_{k}%
})\underset{\eta}{}\underline{h}(Y)],\\
X\underset{g}{}Y  &  =\underline{h}^{-1}[\underline{h}(X)\underset{\eta}%
{}\underline{h}(Y)].
\end{align*}

Hence, eq.(\ref{4.g1}), eq.(\ref{4.g2}), eqs.(\ref{4.g3.1}) and (\ref{4.g3.2}%
), and eq.(\ref{4.g4}) have set the golden formula.
\end{proof}

\section{Metric Adjoint Operators}

Let $g$ be a metric operator on $V,$ i.e., $g\in ext_{1}^{1}(V)$ such that
$g=g^{\dagger}$ and $\det[g]\neq0.$ To each $t\in1$-$ext(\bigwedge
\limits_{1}^{\diamond}V;\bigwedge\limits_{2}^{\diamond}V)$. We define the
metric adjoint operator $t^{\dagger(g)}\in1$-$ext(\bigwedge\limits_{2}%
^{\diamond}V;\bigwedge\limits_{1}^{\diamond}V)$ by
\begin{equation}
t^{\dagger(g)}=\underline{g}^{-1}\circ t^{\dagger}\circ\underline{g}.
\label{4.madj.1}%
\end{equation}

As we can easily see, $t^{\dagger(g)}$ is the unique extensor from
$\bigwedge\limits_{2}^{\diamond}V$ to $\bigwedge\limits_{1}^{\diamond}V$ which
satisfies the following property: for any $X\in\bigwedge\limits_{1}^{\diamond
}V$ and $Y\in\bigwedge\limits_{2}^{\diamond}V$
\begin{equation}
X\underset{g}{\cdot}t^{\dagger(g)}(Y)=t(X)\underset{g}{\cdot}Y.
\label{4.madj.2}%
\end{equation}
This is the `metric version' of the fundamental property given by the formula
$t^{\dagger}(X)\cdot Y=X\cdot t(Y)$ in the paper II of this series.

Finally, we notice the very important formula that
\begin{equation}
\det[t^{\dagger(g)}]=\det[t^{\dagger}]=\det[t]. \label{4.madj.3}%
\end{equation}

\section{Standard Hodge Extensor}

Let ($\{e_{j}\},$ $\{e^{j}\})$ be a pair of $b$-reciprocal bases to each other
for $V,$ i.e., $e_{j}\underset{b}{\cdot}e^{k}=\delta_{j}^{k}.$ Associated to
them we define a non-zero pseudoscalar
\begin{equation}
\tau=\sqrt{e_{\wedge}\cdot e_{\wedge}}e^{\wedge} \label{4.4.1}%
\end{equation}
where $e_{\wedge}\equiv e_{1}\wedge\ldots\wedge e_{n}\in\bigwedge^{n}V$ and
$e^{\wedge}\equiv e^{1}\wedge\ldots\wedge e^{n}\in\bigwedge^{n}V.$ Note that
$e_{\wedge}\cdot e_{\wedge}>0,$ since the scalar product is positive definite.
It will be called a \emph{standard volume pseudoscalar for }$V.$ It has the
fundamental property
\begin{equation}
\tau\cdot\tau=\tau\lrcorner\widetilde{\tau}=\tau\widetilde{\tau}=1,
\label{4.4.2}%
\end{equation}
it follows from the equation $e_{\wedge}\cdot e^{\wedge}=1$.

>From eq.(\ref{4.4.2}), we can get an expansion formula for pseudoscalars
\begin{equation}
I=(I\cdot\tau)\tau. \label{4.4.3}%
\end{equation}

The extensor $\star\in ext(V)$ which is defined by $\star:\bigwedge
V\rightarrow\bigwedge V$ such that
\begin{equation}
\star X=\widetilde{X}\lrcorner\tau=\widetilde{X}\tau, \label{4.4.4}%
\end{equation}
will be called a\emph{\ standard Hodge extensor on }$V.$

It should be noticed that if $X\in\bigwedge^{p}V,$ then $\star X\in
\bigwedge^{n-p}V.$ It means that $\star$ can be also defined as a
$(p,n-p)$-extensor over $V.$

The extensor over $V,$ $\star^{-1}:\bigwedge V\rightarrow\bigwedge V$ such
that
\begin{equation}
\star^{-1}X=\tau\llcorner\widetilde{X}=\tau\widetilde{X} \label{4.4.5}%
\end{equation}
is the \emph{inverse} extensor of $\star.$

Indeed, take $X\in\bigwedge V.$ Eq.(\ref{4.4.2}) gives $\star^{-1}\circ\star
X=\tau\widetilde{\tau}X=X,$ and $\star\circ\star^{-1}X=X\widetilde{\tau}%
\tau=X,$ i.e., $\star^{-1}\circ\star=\star\circ\star^{-1}=i_{\bigwedge V},$
where $i_{\bigwedge V}\in ext(V)$ is the so-called \emph{identity function}
for $\bigwedge V.$

Let us take $X,Y\in\bigwedge V.$ Using the multivector identity $(XA)\cdot
Y=X\cdot(Y\widetilde{A})$ and eq.(\ref{4.4.2}) we get
\begin{equation}
(\star X)\cdot(\star Y)=X\cdot Y. \label{4.4.6}%
\end{equation}
It means that the standard Hodge extensor preserves the euclidean scalar product.

Let us take $X,Y\in\bigwedge^{p}V.$ By using eq.(\ref{4.4.3}) together with
the multivector identity $(X\wedge Y)\cdot Z=Y\cdot(\widetilde{X}\lrcorner
Z),$ and eq.(\ref{4.4.6}) we get
\begin{equation}
X\wedge(\star Y)=(X\cdot Y)\tau. \label{4.4.7}%
\end{equation}
This identity is completely equivalent to the definition of standard Hodge
extensor given by eq.(\ref{4.4.4}).

Take $X\in\bigwedge^{p}V$ and $Y\in\bigwedge^{n-p}V.$ By using the multivector
identity $(X\lrcorner Y)\cdot Z=Y\cdot(\widetilde{X}\wedge Z)$ and
eq.(\ref{4.4.3}) we get
\begin{equation}
((\star X)\cdot Y)\tau=X\wedge Y. \label{4.4.8}%
\end{equation}

\section{Metric Hodge Extensor}

Let $g$ be a metric extensor over $V$ of signature $(p,q),$ i.e., $g\in
ext_{1}^{1}(V)$ such that $g=g^{\dagger}$ and $\det[g]\neq0$. It has $p$
positive and $q$ negative eigenvalues. Associated to a pair ($\{e_{j}\}$,
$\{e^{j}\})$ of $b$-reciprocal bases we can define another non-zero
pseudoscalar
\begin{equation}
\tau_{g}=\sqrt{\left|  e_{\wedge}\underset{g}{\cdot}e_{\wedge}\right|
}e^{\wedge}=\sqrt{\left|  \det[g]\right|  }\tau. \label{4.5.1}%
\end{equation}
It will be called a \emph{metric volume pseudoscalar} for $V.$ It has the
fundamental property
\begin{equation}
\tau_{g}\underset{g^{-1}}{\cdot}\tau_{g}=\tau_{g}\underset{g^{-1}}{\lrcorner
}\tilde{\tau}_{g}=\tau_{g}\underset{g^{-1}}{}\tilde{\tau}_{g}=(-1)^{q}.
\label{4.5.2}%
\end{equation}
It follows from eq.(\ref{4.4.2}) by taking into account the definition of
determinant of a linear operator on $V,$ and recalling that $sgn(\det
[g])=(-1)^{q}.$

An expansion formula for the pseudoscalars can be obtained from
eq.(\ref{4.5.2}), i.e.,
\begin{equation}
I=(-1)^{q}(I\underset{g^{-1}}{\cdot}\tau_{g})\tau_{g}. \label{4.5.3}%
\end{equation}

The extensor $\underset{g}{\star}$ $\in ext(V)$ which is defined by
$\underset{g}{\star}:\bigwedge V\rightarrow\bigwedge V$ such that
\begin{equation}
\underset{g}{\star}X=\widetilde{X}\underset{g^{-1}}{\lrcorner}\tau
_{g}=\widetilde{X}\underset{g^{-1}}{}\tau_{g}, \label{4.5.4}%
\end{equation}
will be called a\emph{\ metric Hodge extensor} on $V.$ It should be noticed
that the definition of $\underset{g}{\star}$ needs the use of both the $g$ and
$g^{-1}$ metric Clifford algebras, a non trivial fact.

It is clear that if $X\in\bigwedge^{p}V,$ then $\underset{g}{\star}%
X\in\bigwedge^{n-p}V.$

The extensor over $V,$ $\underset{g}{\star}^{-1}:\bigwedge V\rightarrow
\bigwedge V$ such that
\begin{equation}
\underset{g}{\star}^{-1}X=(-1)^{q}\tau_{g}\underset{g^{-1}}{\llcorner
}\widetilde{X}=(-1)^{q}\tau_{g}\underset{g^{-1}}{}\widetilde{X}, \label{4.5.5}%
\end{equation}
is the \emph{inverse }extensor of $\underset{g}{\star}.$

Indeed, take $X\in\Lambda V.$ By using eq.(\ref{4.5.2}), we verify that
$\underset{g}{\star}^{-1}\circ\underset{g}{\star}X=(-1)^{q}\tau_{g}%
\underset{g^{-1}}{}\tilde{\tau}_{g}\underset{g^{-1}}{}X=X,$ and $\underset
{g}{\star}\circ\underset{g}{\star}^{-1}X=(-1)^{q}X\underset{g^{-1}}{}%
\tilde{\tau}_{g}\underset{g^{-1}}{}\tau_{g}=X,$ i.e., $\underset{g}{\star
}^{-1}\circ\star=\underset{g}{\star}\circ\underset{g}{\star}^{-1}=i_{\bigwedge
V}.$

Take $X,Y\in\bigwedge V.$ The identity $(X\underset{g^{-1}}{}A)\underset
{g^{-1}}{\cdot}Y=X\underset{g^{-1}}{\cdot}(Y\underset{g^{-1}}{}\widetilde{A})$
and eq.(\ref{4.5.2}) yield
\begin{equation}
(\underset{g}{\star}X)\underset{g^{-1}}{\cdot}(\underset{g}{\star}%
Y)=(-1)^{q}X\underset{g^{-1}}{\cdot}Y. \label{4.5.6}%
\end{equation}

Take $X,Y\in\bigwedge^{p}V.$ Eq.(\ref{4.5.3}), the identity $(X\wedge
Y)\underset{g^{-1}}{\cdot}Z=Y\underset{g^{-1}}{\cdot}(\widetilde{X}%
\underset{g^{-1}}{\lrcorner}Z)$ and eq.(\ref{4.5.6}) allow us to get
\begin{equation}
X\wedge(\underset{g}{\star}Y)=(X\underset{g^{-1}}{\cdot}Y)\tau_{g}.
\label{4.5.7}%
\end{equation}
This remarkable property is completely equivalent to the definition of the
metric Hodge extensor.

Take $X\in\bigwedge^{p}V$ and $Y\in\bigwedge^{n-p}V.$ The use of identity
$(X\underset{g^{-1}}{\lrcorner}Y)\underset{g^{-1}}{\cdot}Z=Y\underset{g^{-1}%
}{\cdot}(\widetilde{X}\wedge Z)$ and eq.(\ref{4.5.3}) yield
\begin{equation}
((\underset{g}{\star}X)\underset{g^{-1}}{\cdot}Y)\tau_{g}=(-1)^{q}X\wedge Y.
\label{4.5.8}%
\end{equation}

It might as well be asked what is the relationship between the standard and
metric Hodge extensors as defined above.

Take $X\in\bigwedge V.$ By using eq.(\ref{4.5.1}), the multivector identity
for an invertible $(1,1)$-extensor $\underline{t}^{-1}(X)\lrcorner
Y=\underline{t}^{\dagger}(X\lrcorner\underline{t}^{*}(Y))$ and the definition
of determinant of a $(1,1)$-extensor, we have
\begin{align*}
\underset{g}{\star}X  &  =\underline{g}^{-1}(\widetilde{X})\lrcorner
\sqrt{\left|  \det[g]\right|  }\tau=\sqrt{\left|  \det[g]\right|  }%
\underline{g}(\widetilde{X}\lrcorner\underline{g}^{-1}(\tau))\\
&  =\frac{\sqrt{\left|  \det[g]\right|  }}{\det[g]}\underline{g}(\widetilde
{X}\lrcorner\tau)=\frac{sgn(\det[g])}{\sqrt{\left|  \det[g]\right|  }%
}\underline{g}\circ\star(X),
\end{align*}
ie.,
\begin{equation}
\underset{g}{\star}=\frac{(-1)^{q}}{\sqrt{\left|  \det[g]\right|  }}%
\underline{g}\circ\star. \label{4.5.9}%
\end{equation}
Eq.(\ref{4.5.9}) is the formula which relates\footnote{It is a very important
formula and good use of it will be done in our theory of the gravitational
field to be presented in another series of papers.} $\underset{g}{\star}$ with
$\star.$

We recall that for any metric operator $g\in ext_{1}^{1}(V)$ there exists a
non-degenerate operator $h\in ext_{1}^{1}(V)$ such that
\begin{equation}
g=h^{\dagger}\circ\eta\circ h, \label{4.5.10}%
\end{equation}
where $\eta\in ext_{1}^{1}(V)$ is an orthogonal metric operator with the same
signature as $g$. Such a $h$ is called a gauge operator for $g.$

The $g$ and $g^{-1}$ metric contracted products $\underset{g}{\lrcorner}$ and
$\underset{g^{-1}}{\lrcorner}$ are related to the $\eta$-metric contracted
product $\underset{\eta}{\lrcorner}$ (recall that $\eta=\eta^{-1}$) by the
following formulas
\begin{align}
\underline{h}(X\underset{g}{\lrcorner}Y)  &  =\underline{h}(X)\underset{\eta
}{\lrcorner}\underline{h}(Y),\label{4.5.11a}\\
\underline{h}^{*}(X\underset{g^{-1}}{\lrcorner}Y)  &  =\underline{h}%
^{*}(X)\underset{\eta}{\lrcorner}\underline{h}^{*}(Y). \label{4.5.11b}%
\end{align}

We can also get a noticeable formula which relates a $g$-metric Hodge extensor
with a $\eta$-metric Hodge extensor.

Now, take $X\in\bigwedge V.$ By using eq.(\ref{4.5.11b}), eq.(\ref{4.5.1}),
the definition of determinant of a $(1,1)$-extensor, eq.(\ref{4.5.10}) and the
equation $\tau_{\eta}=\tau,$ we have
\begin{align*}
\underset{g}{\star}X  &  =\underline{h}^{\dagger}(\underline{h}^{*}%
(\widetilde{X})\underset{\eta}{\lrcorner}\underline{h}^{*}(\underset{g}{\tau
}))=\sqrt{\left|  \det[g]\right|  }\underline{h}^{\dagger}(\underline{h}%
^{*}(\widetilde{X})\underset{\eta}{\lrcorner}\det[h^{*}]\tau)\\
&  =\left|  \det[h]\right|  \det[h^{*}]\underline{h}^{\dagger}(\widetilde
{\underline{h}^{*}(X)}\underset{\eta}{\lrcorner}\underset{\eta}{\tau
})=sgn(\det[h])\underline{h}^{\dagger}\circ\underset{\eta}{\star}%
\circ\underline{h}^{*}(X),
\end{align*}
i.e.,
\begin{equation}
\underset{g}{\star}=sgn(\det[h])\underline{h}^{\dagger}\circ\underset{\eta
}{\star}\circ\underline{h}^{*}. \label{4.5.12}%
\end{equation}
Eq. (\ref{4.5.12}) is the formula which relates $\underset{g}{\star}$ with
$\underset{\eta}{\star}.$

\section{Conclusions}

We showed that any metric Clifford product on $\mathcal{C}\ell(V,g)$ can be
considered as \emph{deformation} of the euclidean Clifford product on
$\mathcal{C}\ell(V),$ induced by the metric extensor $g$. We also proved that
any metric extensor $g$ is decomposable in terms of a \emph{gauge metric
extensor }$h$ and a \emph{fiducial }$b$-\emph{orthogonal extensor }$\eta$
which has the same signature as $g.$ Although $h$ is not unique, since two
$h^{\text{'}}$s satisfying that property differ only by a composition with a
general transformation $\Lambda$ which is a $\eta$-orthogonal $(1,1)$%
-extensor. For the case that $V$ is $4$-dimensional and $\eta$ is a Lorentzian
metric extensor (i.e., with signature $(1,3)$), $\Lambda$ is just a general
Lorentz transformation. The paper contains a proof of the non trivial golden
formula, which as the future papers will show, really deserves its name.
Indeed, the formula is a key in our theory of the intrinsic formulation of
differential geometry on arbitrary manifolds that we will present in future
papers, and also find applications in some some problems of Theoretical
Physics as, e.g., in geometric theories of gravitation and Lagrangian
formulation of the theory of multivector and extensor fields.\medskip

\textbf{Acknowledgement}: V. V. Fern\'{a}ndez is grateful to FAPESP for a
posdoctoral fellowship. W.A. Rodrigues Jr. is grateful to CNPq for a senior
research fellowship (contract 201560/82-8) and to the Department of
Mathematics of the University of Liverpool for the hospitality. Authors are
also grateful to Drs. P. Lounesto, I. Porteous and J. Vaz, Jr. for their
interest in our research and useful discussions.

\end{document}